\documentclass[showpacs,aps,prb,twocolumn,superscriptaddress]{revtex4-1}

\usepackage{graphicx}
\usepackage{hyperref}
\usepackage{epstopdf}
\usepackage{dcolumn}
\usepackage{bm}
\usepackage{amsmath}
\usepackage{amssymb}
\usepackage[dvipsnames]{xcolor}
\usepackage{color} 
\usepackage{ulem}
\usepackage{xspace}
\usepackage{ulem}
 
\newenvironment{psmallmatrix}
  {\left(\begin{smallmatrix}}
  {\end{smallmatrix}\right)}

\usepackage{float}

\renewcommand{\vec}[1]{\boldsymbol{#1}}

\newcommand{\ve}[1]{\boldsymbol{#1}}

\setcitestyle{square,numbers}
\usepackage{here}

\begin{document}

\title{Negative sign free formulations of generalized Kitaev models with higher symmetries}

\author{Toshihiro Sato}
\affiliation{\mbox{Institut f\"ur Theoretische Physik und Astrophysik, Universit\"at W\"urzburg, 97074 W\"urzburg, Germany}}
\author{Fakher F. Assaad}
\affiliation{\mbox{Institut f\"ur Theoretische Physik und Astrophysik, Universit\"at W\"urzburg, 97074 W\"urzburg, Germany}}
\affiliation{\mbox{W\"urzburg-Dresden Cluster of Excellence ct.qmat, Am Hubland, 97074 W\"urzburg, Germany}}

\begin{abstract}
We provide a negative-sign-free formulation of the auxiliary field quantum Monte Carlo algorithm for generalized Kitaev models with higher symmetries.
Our formulation is based on the Abrikosov fermion representation of the spin-1/2 degree of freedom and the phase pinning approach  [Phys. Rev. B~{\bf 104}, L081106 (2021)]. 
Enhancing the number of fermion flavors or orbitals from one to $N$ allows one to generalize the inherent $Z_2$ global symmetry to Z$_2$$\times$SU($N$)$_o$. 
Using this general approach, we study the Z$_2$$\times$SU($2$)$_o$ Kitaev-Heisenberg model reflecting the competition between the isotropic Heisenberg exchange and Kitaev-type bond-directional exchange interactions.
We show that the symmetry enhancement provides a path to escape frustration and  that  the spin liquid phases  in the original Z$_2$ symmetric model are not present in this model.
Nevertheless, the ground-state phase diagram is extremely rich and has points with higher global and local continuous symmetries  as  well  as   de-confined  quantum  critical points. 
\end{abstract}

\maketitle

\section{Introduction}
Quantum Monte Carlo (QMC) methods play an important role in the discovery of many fascinating states of correlated quantum matter.
With this approach one can numerically solve target models for a given lattice size and temperature without any  further   approximations. In particular, it excels at computing thermodynamic properties.
However,  many spin and fermion models suffer from  the infamous negative sign problem that renders the computational cost exponential in  the  volume of the system $V$ and in  the  inverse temperature $\beta$ \cite{Troyer05}.
The severity of the negative sign problem depends on  model  parameters  and  on the  specific  formulation.   In  some  cases,  one  can  use  symmetry  arguments  to  
avoid  it  altogether  \cite{Wu04,Wei16,Li16}.  It   most  cases,  the  sign  problem  remains and  optimization strategies to alleviate it  can  be  put  forward  \cite{Ulybyshev19,Ulybyshev19a,Wan20,Hangleiter20}.

In the past years, there has been sustained progress in defining classes of models that are free of the negative sign problem in the realm of the auxiliary field QMC (AFQMC) algorithm.
The AFQMC methods for fermions~\cite{Blankenbecler81,White89,ALF_v1} that we will consider here are based on a Trotter decomposition and Hubbard-Stratonovich transformation of  the interaction.   The partition function can then generically be represented as 
\begin{equation}
\label{eq:ZAFQMC}
    Z  = {\rm{Tr}} \left[e^{-\beta \hat{H}} \right] = \int  d \Phi(i,\tau)     e^{-S  \left( \Phi(i,\tau)   \right) },
\end{equation}
where $\Phi$ corresponds to  a  space ($i$) and time ($\tau$) dependent Hubbard-Stratonovich field.
$S$ is the action  of a single particle Hamiltonian subject to the field $\Phi$  and is generically  given by
\begin{equation}
\label{eq:action}
S(\Phi)    =  S_0(\Phi) - \log   \text{Tr} \left[ {\cal T}   e^{-\int_{0}^{\beta}  d \tau   \sum_{i ,j }\hat{c}^{\dagger}_{i}  h_{i,j}(\tau)   \hat{c}^{\phantom\dagger}_{i} } \right]
\end{equation}
with a real bosonic action $S_0$ and a single body Hamiltonian with $\Phi$ and time $\tau$ dependent matrix $h_{i,j}(\tau)  $.
The fermion operator $\hat{c}^{\dagger}_{i}$  creates a particle in  the single  particle  state labeled by $i$.
The trace over the fermion degrees of freedom is complex number,  thus leading to $ \text{Im}  S   \in [0, 2 \pi ]$.  
Since the MC importance sampling of the field $\Phi$ is implemented by a weight function $ |e^{-S(\Phi)} | $,  the average sign is given as the reweighting factor $\langle   \text{sign}  \rangle   =  \int d \Phi e^{- S (\Phi )}  /  \int d \Phi  | e^{- S (\Phi )} |   $.   
It has been shown  that using symmetry based strategies \cite{Wu04,Wei16,Li16},   $\text{Im} S$   can be pinned  to zero  thereby   defining  sign free, $\langle   \text{sign}  \rangle=1$,   models.  
For instance,  in Ref.~\cite{Li16},  the negative sign problem is absent if one can find two anti-unitary operators  that mutually anti-commute and that commute with  the aforementioned $h_{i,j} $ matrix  for each field configuration.   
This symmetry based strategy has led to an ever  growing class of negative sign free model Hamiltonians~\cite{Huffman14,Schattner15,SatoT17,SatoT17_1,Liu18,Ippoliti18,WangZ20,Pan20} that can be simulated   with the AFQMC approach.

It is natural to  ask how to optimize the negative sign problem in the absence of negative-sign-free formulations.
The idea  that  we  will follow is  that reducing the fluctuations of  $\text{Im} S$ will reduce  the severity of the negative sign problem.
In particular if one can design a formulation of the path integral such that there exits one anti-unitary operator that commutes with $ h_{\ve{i},\ve{j}} $, then the phase is pinned to $\text{Im}S = 0,\pi$. 
In a recent publication~\cite{SatoT21_1}, we have   achieved  this  for a large class of  frustrated spin models.  This includes   the generalized Kitaev model 
for  which  this phase pinning strategy to mitigate the severity of the negative sign problem  opens a window of temperatures relevant to experiments where QMC simulations can be carried out. 
A further important consequence of this phase quantization, ${\rm{Im}} S = 0, \pi $, is that it allows us to define a set of models with higher symmetries that are free of the negative sign problem. 
In particular, and as we will see below,  attaching an additional orbital  index, $n$,  to the   fermion  operator  with  $ n = 1 \cdots N$, and $N$ even, allows to avoid the negative sign problem.  
In this paper, we will use the phase pinning strategy to provide negative-sign-free formulations of   Z$_2$$\times$SU($N$)$_o$ generalized Kitaev models.  
Our  motivation is  to   investigate  if  this  line of   negative sign free model  building provide  interesting  phase  diagrams.     We note  that such  ideas  have  already been put  forward in the context of the doped Hubbard model \cite{Assaad02a}.

This paper is organized as follows.  
We start in Sec.~\ref{sec:LargeNgeneralizations} by  providing a demonstration of the phase pinning approach, and then show how to implement this idea for  Z$_2$$\times$SU($N$)$_o$ generalized Kitaev models.
In Sec.~\ref{sec:Results}, we use this approach to explore the ground-state properties of the Z$_2$$\times$SU($2$)$_o$ Kitaev-Heisenberg model.
Section ~\ref{sec:Summary} concludes this paper with a summary  and  a  discussion of the pros and cons of  such an approach. 

\section{AFQMC formulations of generalized Kitaev models with higher symmetries}
\label{sec:LargeNgeneralizations}

In this section we detail the formulation of the auxiliary field quantum Monte Carlo (AFQMC) algorithm for generalized Kitaev models with higher symmetries. 
The original model Hamiltonian with  inherent Z$_2$ global symmetry reads
\begin{eqnarray}
\label{Eq:KHM}
\hat{H}   =  \sum_{\ve{i},\ve{j},\alpha,\beta}  \Gamma_{\ve{i},\ve{j}}^{\alpha,\beta} \hat{S}_{i}^{\alpha}  \hat{S}_{\ve{j}}^{\beta}  + \sum_{\ve{i},\ve{j}}  J_{\ve{i},\ve{j}}  \hat{\ve{S}}_{\ve{i}}  \cdot \hat{\ve{S}}_{\ve{j}}.
\end{eqnarray}
Here the spin-1/2 degree of freedom $\hat{S}_{\ve{i}}^{\alpha}$ with $\alpha=(x,y,z)$ resides on a graph with sites labeled by $\ve{i},\ve{j}$ of the honeycomb lattice.
While $\Gamma_{\ve{i},\ve{j}}^{\alpha,\beta}$ defines the potentially frustrated spin model, $J_{\ve{i},\ve{j}}$ accounts for non-frustrating exchange couplings.
To formulate the algorithm we represent the spin-1/2 degree of freedom in terms of Abrikosov fermions
\begin{eqnarray}
\label{Eq:AFrep}
\hat {\ve{S}}_{\ve{i}} = \frac{1}{2} \hat{\ve{ f}}^{\dagger}_{\ve{i}}   \hat{ \ve{\sigma}}  \hat{ \ve{f}} ^{\phantom\dagger}_{\ve{i}}
\end{eqnarray}
where $\hat{\ve{f}}^{\dagger}_{\ve{i}}  \equiv  \left(\hat {f}^{\dagger}_{\ve{i},\uparrow}, \hat f^{\dagger}_{\ve{i},\downarrow} \right) $ is a two-component fermion on site $\ve{i}$ with constraint $\hat{\ve{f}}^{\dagger}_{\ve{i}}  \hat{\ve{f}}^{\phantom\dagger}_{\ve{i}}   = 1$ and $\ve{\sigma}$ corresponds to the vector of 
Pauli spin-1/2 matrices.
We  work in an unconstrained  Hilbert space  and  enforce  the  constraint $\hat{\ve{f}}^{\dagger}_{\ve{i}}  \hat{\ve{f}}^{\phantom\dagger}_{\ve{i}}   = 1$ by 
including  a  Hubbard-$U$ term on each site. 
Similar ideas were used in the framework  of Kondo lattice models~\cite{Capponi00,Assaad99a,SatoT17_1}.
In a recent publication, Ref.~\cite{SatoT21_1}, we introduced  the phase pinning  idea  in the realm of the AFQMC
 for the original Z$_2$ symmetric model of Eq.~(\ref{Eq:KHM}).
The key technical insight is that if one can design a formulation of the path integral such that there exits one anti-unitary operator that commutes with the one-body Hamiltonian coupled  to the auxiliary field, then the imaginary part  of the action $S$ is pinned to
 \begin{equation}
 \label{Eq:phasepinning}
	\text{Im}S = 0, \pi.
\end{equation} 

In the fermion representation and  so-called phase pinning approach, the model of Eq.~(\ref{Eq:KHM})  can be simulated using
\begin{eqnarray}
\label{Eq:HQMC}
\hat{H}_{{\rm QMC}}    &= & \sum_{ \ve{i} ,\ve{j}, \alpha, \beta }\frac{|\Gamma_{\ve{i},\ve{j}}^{\alpha,\beta} |}{2}  \left( \hat{S}_{\ve{i}}^{\alpha}  + \frac{\Gamma_{\ve{i},\ve{j}}^{\alpha,\beta}} {|\Gamma_{\ve{i},\ve{j}}^{\alpha,\beta} |}   \hat{S}_{\ve{j}}^{\beta}  \right)^2  \nonumber \\
& & - \sum_{ \ve{i} ,\ve{j}}  \frac{J_{\ve{i},\ve{j}}}{8} \left[  \left(  \hat{D}^{\dagger}_{\ve{i},\ve{j}}  + \hat{D}^{\phantom \dagger}_{\ve{i},\ve{j}}  \right)^2+ \left(i \hat{D}^{\dagger}_{\ve{i},\ve{j}}   -i \hat{D}^{\phantom\dagger}_{\ve{i},\ve{j}}  \right)^2  \right] \nonumber \\
& &+U  \sum_{\ve{i}} \left( \hat{\ve{f}}^{\dagger}_{\ve{i}} \hat{\ve{f}}^{\phantom\dagger}_{\ve{i}} - 1 \right)^2
\end{eqnarray}
with $ \hat{D}^{\dagger}_{\ve{i},\ve{j}} = \hat{\ve{f}}^{\dagger}_{\ve{i}} \hat{\ve{f}}^{\phantom\dagger}_{\ve{j}} $.     
Here, $\left(  \hat{\ve{f}}^{\dagger}_{\ve{i}}  \hat{\ve{f}}^{\phantom\dagger}_{\ve{i}}  -1  \right)^2$ commutes with $\hat H_{\rm{QMC}}$ such that the  $\hat{\ve{f}}$-fermion parity  $ (-1)^{\hat{\ve{f}}^{\dagger}_{\ve{i}} \hat{\ve{f}}^{\phantom\dagger}_{\ve{i}} }  $ is a local conserved quantity.
Owing to this symmetry property, the additional Hubbard $U$ term with $U>0$ will project very efficiently on the odd parity sector ${(-1)^{ \hat{\ve{f}}^{\dagger}_{\ve{i}}  \hat{\ve{f}}^{\phantom\dagger}_{\ve{i}} }   = -1 }$, thus imposing the constraint $\hat{\ve{f}}^{\dagger}_{\ve{i}}  \hat{\ve{f}}^{\phantom\dagger}_{\ve{i}}   = 1$.
In this sector, $ \left. \hat{H}_{\rm{QMC}} \right|_{(-1)^{ \hat{\ve{f}}^{\dagger}_{\ve{i}}  \hat{\ve{f}}^{\phantom\dagger}_{\ve{i}} }   = -1 }= \hat{H} + C $ where $C$ is a constant.
In Eq.~(\ref{Eq:HQMC})  the interaction is a sum of  perfect squares   and can hence  be  directly    implemented in the ALF (Algorithms for Lattice Fermions)~\cite{ALF_v1,ALF_v2}  formulation of the AFQMC algorithm~\cite{Blankenbecler81,White89,Assaad08_rev}.
Since we  assume  the $J_{\ve{i},\ve{j}}$ exchange couplings  to be non-frustrating, one can find a set of Ising spins $s_{\ve{i}}= \pm 1$  such that for each bond with $J_{\ve{i},\ve{j}}  \neq 0 $, $J_{\ve{i},\ve{j}}  = |  J_{\ve{i},\ve{j}} | \left(- s_{\ve{i}} s_{\ve{j}} \right)$.
After a Trotter decomposition and Hubbard-Stratonovich transformation the partition function can be written as
 \begin{eqnarray}
 \label{ZN1}
    Z &=& 
    {\rm{Tr}} \left[ e^{- \beta \hat{H}_{\rm{QMC}}}\right]      \nonumber \\
      &\propto  & \int   D \left\{ \chi_{\ve{i},\ve{j}}^{\alpha,\beta}(\tau) , {\rm{Re}}Z_{\ve{i},\ve{j}}(\tau),  {\rm{Im}}Z_{\ve{i},\ve{j}}(\tau) , \lambda_{\ve{i}}(\tau) \right\}      \nonumber \\
      & & \times 
      e^{- S\left( \left\{  \chi_{\ve{i},\ve{j}}^{\alpha,\beta}(\tau), Z_{\ve{i},\ve{j}}(\tau), \lambda_{\ve{i}}(\tau)   \right\} \right)}
 \end{eqnarray}
with an inverse temperature $\beta$ and an imaginary time $\tau$. 
The action in given field configuration  $\chi_{\ve{i},\ve{j}}^{\alpha,\beta}(\tau)$, $\lambda_{\ve{i}}(\tau)  \in \mathbb{R} $  and $Z_{\ve{i},\ve{j}}(\tau)  \in \mathbb{C} $, corresponds to
 \begin{eqnarray}	
S \left( \left\{ \chi,Z,\lambda \right\}  \right) =   && 
     \int_{0}^{\beta} d \tau  \left[  \sum_{\ve{i},\ve{j}, \alpha,\beta}  \frac{ \left( \chi_{\ve{i},\ve{j}}^{\alpha,\beta}(\tau)\right)^2 }{|\Gamma_{\ve{i},\ve{j}}^{\alpha,\beta} |}     \right.  \nonumber \\
    &&  
    \left. + \sum_{\ve{i},\ve{j}} \frac{ |   Z_{\ve{i},\ve{j}}(\tau)|^2 } { 4 |J_{\ve{i},\ve{j}} | }  + \sum_{i} \frac{ \lambda_{\ve{i}}(\tau)^2 }{ 2U } \right]   \nonumber  \\  & &    -  \ln {\rm{Tr}}  {\cal T} e^{-\int_{0}^{\beta } d \tau \hat{h}(   \left\{ \chi,Z,\lambda \right\})}
\end{eqnarray}
with 
\begin{eqnarray}
	\hat{h} (   \left\{ \chi,Z,\lambda \right\} )    &= & \sum_{\ve{i},\ve{j}, \alpha, \beta  }  i  \chi_{\ve{i},\ve{j}}^{\alpha,\beta}(\tau) \left( \hat{S}_{\ve{i}}^{\alpha}  + \frac{\Gamma_{\ve{i},\ve{j}}^{\alpha,\beta}} {|\Gamma_{\ve{i},\ve{j}}^{\alpha,\beta} |}   \hat{S}_{\ve{j}}^{\beta}  \right)   \nonumber \\
& &  + \sum_{\ve{i},\ve{j}, \delta } \sqrt{-s_{\ve{i}} s_{\ve{j}}} \left(   Z_{\ve{i},\ve{j}} (\tau)  \hat{D}^{\dagger}_{\ve{i},\ve{j}}  + \overline{Z_{\ve{i},\ve{j}} (\tau) } \hat{D}^{\phantom \dagger}_{\ve{i},\ve{j}}  \right)  \nonumber \\
& &
+ \sum_{\ve{i}} i \lambda_{\ve{i}}(\tau) \left( \hat{\ve{f}}^{\dagger}_{\ve{i}} \hat{\ve{f}}^{\phantom\dagger}_{\ve{i}}  - 1 \right). 
\end{eqnarray}
In the above, the first sum runs over bonds and spin indices with $ \Gamma_{\ve{i},\ve{j}}^{\alpha,\beta} \neq 0$ and the second sum over bonds with $J_{\ve{i},\ve{j}}\neq 0$.
Now consider the anti-unitary transformation  
\begin{equation}
\label{Symm.eq2}
	  \hat{T} \alpha \hat{f}^{\dagger}_{\ve{i},\sigma} \hat{T}^{-1} =  \overline{\alpha}  s_i \hat{f}^{\phantom\dagger}_{\ve{i},\sigma}~~,~~\alpha  \in \mathbb{C}
\end{equation}
such that one will show that
\begin{equation}	
	 \hat{T}  \hat{h} (   \left\{ \chi,Z,\lambda \right\} )   \hat{T}^{-1}  =   \hat{h} (   \left\{ \chi,Z,\lambda \right\} ).  
\end{equation}
Hence, one can find the anti-unitary symmetry under which $\hat{h}$ is invariant, thus satisfying
\begin{equation}
\label{Eq:Quant}
	   {\rm{Im}} S  \left( \left\{ \chi,Z,\lambda \right\}  \right)     =  0, \pi.
\end{equation}

The important consequence of the aforementioned phase pinning approach is that it allows to define a set of models with higher symmetries that are free of the negative sign problem.
Now consider the following partition function
 \begin{eqnarray}
 \label{ZN2}
    Z_N &=& \int   D \left\{ \chi_{\ve{i},\ve{j}}^{\alpha,\beta}(\tau) , {\rm{Re}}Z_{\ve{i},\ve{j}}(\tau),  {\rm{Im}}Z_{\ve{i},\ve{j}}(\tau) , \lambda_{\ve{i}}(\tau) \right\}      \nonumber \\
      & & \times 
      e^{- NS\left( \left\{  \chi_{\ve{i},\ve{j}}^{\alpha,\beta}(\tau), Z_{\ve{i},\ve{j}}(\tau), \lambda_{\ve{i}}(\tau)   \right\} \right)}
 \end{eqnarray}
which owing to Eq.~(\ref{Eq:Quant}) is free of the negative sign problem at even values of $N$. 
$Z_N$  corresponds to the partition function of the Hamiltonian in Eq.~(\ref{Eq:HQMC}) where the fermion operator, $\hat{\ve{f}}$, acquires an additional orbital index $n$ running from $1,...,N$  reflecting  the  SU($N$)$_o$ global symmetry.
Enhancing the number of fermion orbitals from one to $N$ allows one to generalize the inherent Z$_2$ global symmetry in the generalized Kitaev model in Eq.~(\ref{Eq:KHM}) to a set of Z$_2$$\times$ SU($N$)$_o$.
The Hamiltonian that we will simulate reads
\begin{eqnarray}
\label{Eq:HQMCN}
\hat{H}_{{\rm QMC}} &= & \sum_{\ve{i},\ve{j}, \alpha, \beta }\frac{|\Gamma_{\ve{i},\ve{j}}^{\alpha,\beta} |}{2}  \left( \hat{S}_{i}^{\alpha}  + \frac{\Gamma_{\ve{i},\ve{j}}^{\alpha,\beta}} {|\Gamma_{\ve{i},\ve{j}}^{\alpha,\beta} |}   \hat{S}_{j}^{\beta}  \right)^2  \nonumber \\
& & - \sum_{ \ve{i},\ve{j}}  \frac{J_{\ve{i},\ve{j}}}{8} \left[  \left(  \hat{D}^{\dagger}_{\ve{i},\ve{j}}  + \hat{D}^{\phantom \dagger}_{\ve{i},\ve{j}}  \right)^2+ \left(i \hat{D}^{\dagger}_{\ve{i},\ve{j}} -i \hat{D}^{\phantom\dagger}_{\ve{i},\ve{j}}  \right)^2  \right] \nonumber \\
& &+U  \sum_{\ve{i}} \left[\sum_{s, n} \left(\hat{f}^{\dagger}_{\ve{i},s,n} \hat{f}^{\phantom\dagger}_{\ve{i},s,n}  - 1/2 \right) \right]^2.
\end{eqnarray}
In the above,
\begin{eqnarray}
\label{Eq:DN}
 \hat{D}^{\dagger}_{\ve{i},\ve{j}} = \sum_{s,n}\hat{f}^{\dagger}_{\ve{i},s,n}  \hat{f}^{\phantom\dagger}_{\ve{j},s,n}
\end{eqnarray}
and 
\begin{eqnarray}
\label{Eq:SN}
	\hat{S}_{\ve{i}}^{\alpha} =\frac{1}{2} \sum_{n,s,s'}\hat{f}^{\dagger}_{\ve{i},s,n} \sigma^{\alpha}_{s,s' }   \hat{f}^{\phantom\dagger}_{\ve{i},s',n}
\end{eqnarray}
with the local constraint $\sum_{s,n}\hat{f}^{\dagger}_{\ve{i},s,n}  \hat{f}^{\phantom\dagger}_{\ve{i},s,n}=N$.
$\hat{S}_{\ve{i}}^{\alpha}$   satisfy   the commutation relations
\begin{eqnarray}
\label{Eq:SN-cr}
[\hat{S}_{\ve{i}}^{\alpha}, \hat{S}_{\ve{i}}^{\beta}]=i \epsilon_{\alpha,\beta,\gamma}\hat{S}_{\ve{i}}^{\gamma}
\end{eqnarray}
such that the  $SU(2)_s$ spin algebra is still valid.    While  $s$- corresponds to a spin index,  we  will refer to  $n$  in terms of  
an  orbital index.   
The generators of SU($N$)$_o$ are
\begin{eqnarray}
 \label{TOP}
\hat{T}_{\ve{i}}^{\alpha}=   \sum_{s,n,n'} \hat{f}^{\dagger}_{\ve{i},s,n}T_{n,n'}^{\alpha}\hat{f}^{\phantom\dagger}_{\ve{i},s,n'}
\end{eqnarray}   
that we  choose  to  satisfy  with the  normalization  condition
\begin{equation}
	   \text{Tr} \left[   \hat{T}^{\alpha},  \hat{T}^{\beta} \right]    =  \frac{1}{2} \delta_{\alpha,\beta}. 
\end{equation}
Thereby  at  $N=2$,   $\hat{T}^{\beta}  =  \frac{1}{2} \hat{\tau}^{\beta}$   with   the Pauli spin matrices $\ve{\hat{\tau}}   =  \left[  
\begin{psmallmatrix}0 & 1\\1 & 0\end{psmallmatrix}  , 
\begin{psmallmatrix}0 & -i\\i & 0\end{psmallmatrix} , 
\begin{psmallmatrix}1 & 0 \\0  & -1\end{psmallmatrix} 
\right]$. 
Since  we  use  a  fermionic   representation and  will  impose  the  constraint of  $N$ particles  on the $2N$  orbitals   of  each unit  cell,  
the    representation of SU($N$)$_o$  we  consider corresponds  to the  totally  antisymmetry self-adjoint  one. 
By  construction, rotations  in spin and in orbital space   commute
\begin{equation}
   \left[   \hat{T}_{\ve{i}}^{\alpha},  \hat{S}_{\ve{j}}^{\beta}  \right]   = 0. 
\end{equation}

\begin{figure}[t]
\centerline{\includegraphics[width=0.45\textwidth]{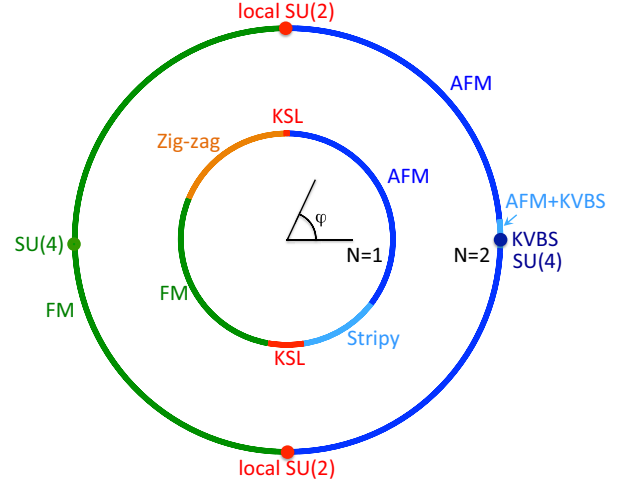}}
\caption{
The ground-state phase diagram of the Z$_2$$\times$ SU($N$)$_o$ Kitaev-Heisenberg model as a function of the angle $\varphi$.}  
\label{fig:pd}   
\end{figure}

\section{Results}
\label{sec:Results}

For concreteness,  we consider on a link $\ve{i},\ve{j}$ defining a nearest-neighbor $\delta$-bond of the honeycomb lattice, $ \Gamma_{\delta}^{\alpha,\beta } =  2K \delta_{\alpha,\beta} \delta_{\delta,\alpha} $ and   $J_{\delta} = J $ in Eq.~(\ref{Eq:HQMCN})  to simulate the Kitaev-Heisenberg model at $N=2$. 
When $K$ is set to zero, the global symmetry inherent in this Heisenberg model is SU(4).
At any finite values of $K$, this symmetry is reduced to a Z$_2$$\times$SU(2)$_o$ one in which Z$_2$ corresponds to the invariance under the inversion $  \ve{\ve{S}}_{\ve{i}}  \rightarrow -\ve{\ve{S}}_{\ve{i}}$.
We adopt the parametrization  $K=A{\rm{sin}}(\varphi)$,  $J=A{\rm{cos}}(\varphi)$,  with  $A=\sqrt{K^2+J^2}$. 
We simulated lattices with $L \times L$ unit cells (each containing two orbitals, i.e., $N_s=2L^2$ sites on the honeycomb lattice) and periodic boundary conditions.
Henceforth, we use $A=1$ as the energy unit.
As for the Trotter discretization  we have used $\Delta\tau=0.1$ and values of $\beta U =10$  were found to be sufficient to guarantee projection to the odd  parity sector.
We have used a range of temperature $T  \in [1/200, 1/80]$  depending upon the considered parameter and this choice of temperature yields results representative of the ground state.
Fig.~\ref{fig:pd} shows the ground-state phase diagram as a function of the angle $\varphi$ as obtained from  a finite-size scaling analysis.
To map out the phase diagram, we measure correlation functions of the spin operators $\hat{\ve{S}}_{\ve{i}}$, SU($N$)$_o$ generators $\hat{\ve{T}}_{\ve{i}}$, and dimer operators $\hat{D}_{\ve{i},\alpha}^{T}=\hat{\vec{T}}_{\ve{i}}\cdot \hat{\vec{T}}_{\ve{i}+\ve{\delta}_\alpha}$ for each-$\delta_{\alpha}$ bond.
The ground-state phase diagram at $N=1$ has been studied to date~\cite{Chaloupka-2013,Gohlke-2017} [see Fig.~\ref{fig:pd}], and leads to antiferromagnetic (AFM), ferromagnetic (FM),  zig-zag, and stripy ordered states, and Kitaev spin liquid (KSL) states.
We find that the $N=2$ deformations of the $N=1$ Kitaev-Heisenberg model  show  different ground-state properties.
Aside from the AFM and FM ordered states with  spontaneous broken SU(2)$_o$ symmetry and the Kekul\'e valence bond solid (KVBS) ordered state with a spontaneously broken translation symmetry, we observe that the KSL states are absent and that states with higher global SU(4) and local SU(2)$_o$ continuous symmetries arise.

\begin{figure}[t]
\centerline{\includegraphics[width=0.48\textwidth]{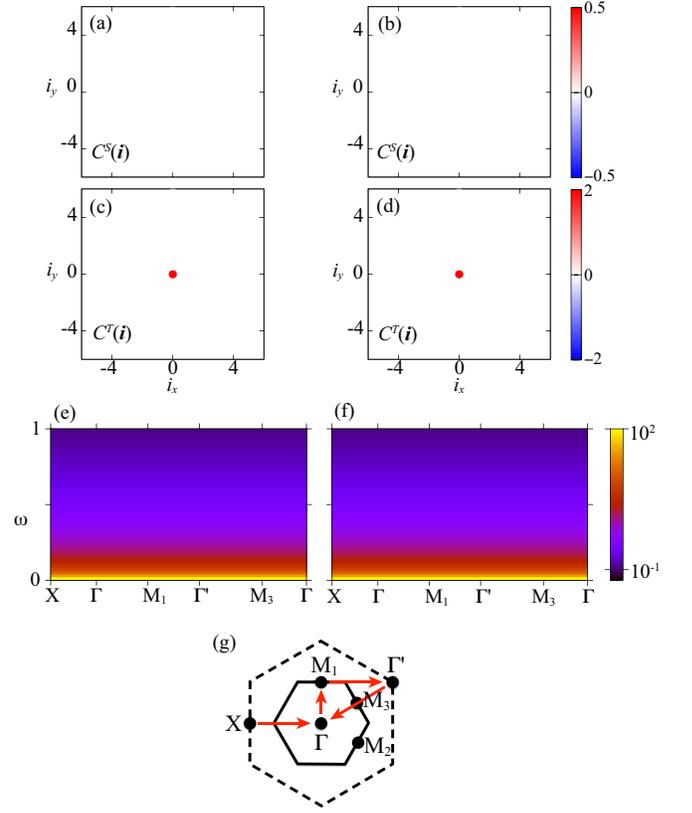}}
\caption{ 
Real-space correlations of $\hat{S}^{1}$  [(a), (b)] and $\hat{\vec{T}}$ [(c), (d)] and dynamical structure factor of $\hat{\vec{T}}$ generators, $C^T(\boldsymbol{q}, \omega)$ [(e), (f)] in the Kitaev limits at $N=2$. (a), (c), (e)  $\varphi/\pi=0.5$ [AFM Kitaev limit] and (b), (d), (f) $\varphi/\pi=1.5$ [FM Kitaev limit]. (g) First (solid) and second (dashed line) Brillouin zones. Results in (e), (f) correspond to scans along the red line. Here, $L=9$ and $T=1/200$.
}
\label{fig:CrKitaev}   
\end{figure}

\begin{figure*}[t]
\centerline{\includegraphics[width=0.95\textwidth]{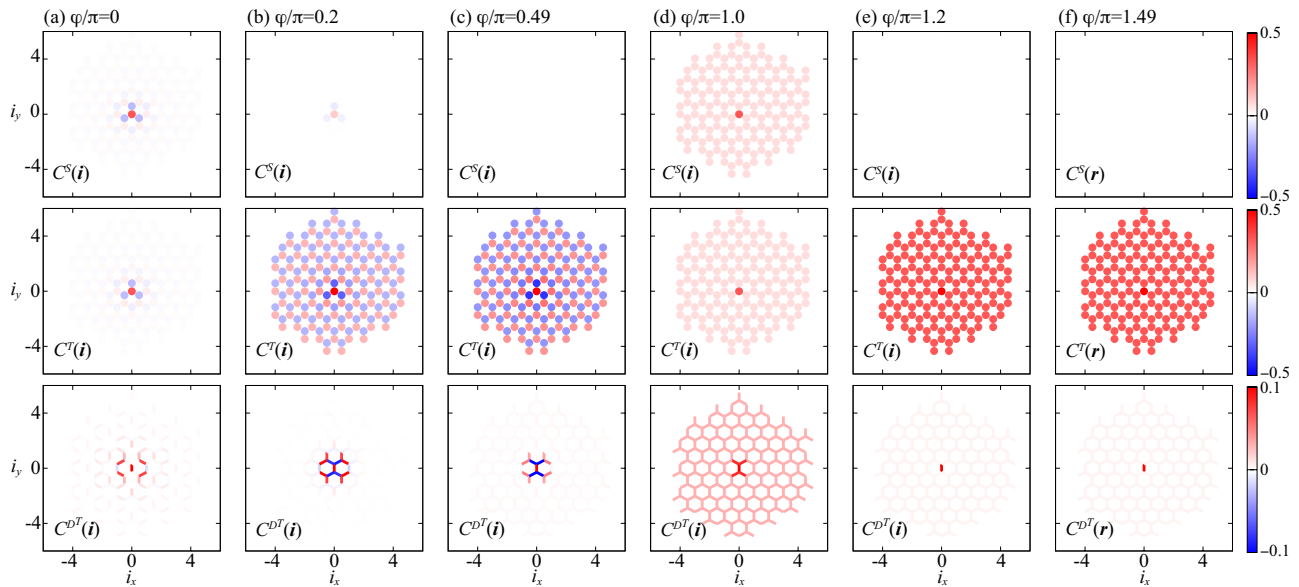}}
\caption{ 
Real-space correlations of $\hat{S}^{1}$ (top panel),  $\hat{\vec{T}}$ (middle panel), and $\hat{D}^{T}$ (bottom panel) at different values of the angle $\varphi$. (a), (b), (d), (e) $T=1/80$ and (c), (f) $T=1/200$. Here, $L=9$ and $N=2$. }
\label{fig:Crangle}   
\end{figure*}

\subsection{Kitaev limits}
We first examine the Kitaev limits (AFM case, $\varphi/\pi=0.5$, and FM case, $\varphi/\pi=1.5$), the Z$_2$$\times$ SU($N$)$_o$ Kitaev model at $N=2$.    In this  case  the  Hamiltonian is  given by: 
\begin{eqnarray}
\label{Eq:S1GS-3}
\hat{H} =  | K |  \sum_{b}   \hat{O}_{b}^{\dagger}\hat{O}^{\phantom\dagger}_{b} 
+U  \sum_{\ve{i}} \left(  \hat{n}_{\ve{i}} -    N   \right)^2
\end{eqnarray}
with $\hat{O}_{b}=\hat{S}_{\ve{i}}^{\alpha}  + \frac{K} {|K|}   \hat{S}_{\ve{i}+\ve{\delta}_{\alpha}}^{\alpha}$ and  
$  \hat{n}_{\ve{i}} =   \sum_{s, n}  \hat{f}^{\dagger}_{\ve{i},s,n} \hat{f}^{\phantom\dagger}_{\ve{i},s,n} $.
In  the  large U-limit,  charge fluctuations are  suppressed  such  that  $ \hat{n}_{\ve{i}}   |\Psi_0 \rangle  =  N  |\Psi_0 \rangle  $  and  
\begin{equation}
	E_0   =  | K |   \sum_{b}   || \hat{O}^{\phantom\dagger}_{b}  | \Psi_0 \rangle ||^2  \geq  0. 
\end{equation}
As  a consequence, any  wave  function   that  satisfies  $  \langle  \Psi_0 |  \hat{H}| \Psi_0 \rangle = 0 $  is  degenerate  with the 
ground  state.   We  now   show  that   the  ground state   degeneracy  is  at  least  $3^{N_s}$    where $N_s$  is  the number of sites 
on the  honeycomb lattice. 
At $N=2$    each  site   hosts  six  states  that we  can conveniently classify  as  spin-singlets  and  orbital-triplets  
\begin{eqnarray} 
\label{Eq:S1GS-1}
|1,1 \rangle_{o,\ve{i}}&=&\hat{f}^{\dagger}_{\ve{i},\uparrow,1}\hat{f}^{\dagger}_{\ve{i},\downarrow,1}|0 \rangle,\nonumber \\
 |1,-1 \rangle_{o,\ve{i}}&=&\hat{f}^{\dagger}_{\ve{i},\uparrow,2}\hat{f}^{\dagger}_{\ve{i},\downarrow,2}|0 \rangle, \nonumber \\
 |1,0 \rangle_{o,\ve{i}}&=&\frac{1}{\sqrt{2}} (\hat{f}^{\dagger}_{\ve{i},\uparrow,1}\hat{f}^{\dagger}_{\ve{i},\downarrow,2}+\hat{f}^{\dagger}_{\ve{i},\uparrow,2}\hat{f}^{\dagger}_{\ve{i},\downarrow,1})|0\rangle
\end{eqnarray}    
for   which 
\begin{eqnarray}
\label{eq:S1_orb}
	   \hat{S}^{\alpha}_{\ve{i}} |1, m \rangle_{o,\ve{i}}   &=& 0,     \nonumber  \\ 
	  \hat{T}^{z}_{\ve{i}} |1, m \rangle_{o,i} &=& m  |1, m \rangle_{o,\ve{i}}, \nonumber  \\   
	 \left( \hat{\ve{T}}_{\ve{i}} \right)^2  |1, m \rangle_{o,\ve{i}} &=& 1(1+1) |1, m \rangle_{o,\ve{i}} 
\end{eqnarray}
  as   well  as   orbital-singlets  and    spin-triplets  
\begin{eqnarray} 
\label{Eq:S1GS-1}
 |1,1 \rangle_{s,\ve{i}}&=&\hat{f}^{\dagger}_{\ve{i},\uparrow,1}\hat{f}^{\dagger}_{\ve{i},\uparrow,2}|0 \rangle,\nonumber \\
|1,-1 \rangle_{s,\ve{i}}&=&\hat{f}^{\dagger}_{\ve{i},\downarrow,1}\hat{f}^{\dagger}_{\ve{i},\downarrow,2}|0 \rangle, \nonumber \\
 |1,0 \rangle_{s,\ve{i}}&=&\frac{1}{\sqrt{2}} (\hat{f}^{\dagger}_{\ve{i},\uparrow,1}\hat{f}^{\dagger}_{\ve{i},\downarrow,2}+ 
    \hat{f}^{\dagger}_{\ve{i},\downarrow,1}  \hat{f}^{\dagger}_{\ve{i},\uparrow,2})|0\rangle
\end{eqnarray}  
for   which 
\begin{eqnarray}
	   \hat{T}^{\alpha}_{\ve{i}} |1, m \rangle_{s,\ve{i}}   &=& 0     \nonumber  \\ 
	  \hat{S}^{z}_{\ve{i}} |1, m \rangle_{s,\ve{i}}  &=&  m  |1, m \rangle_{s,\ve{i}}, \nonumber  \\ 
	  \left( \hat{\ve{S}}_{\ve{i}} \right)^2  |1, m \rangle_{s,\ve{i}} &=& 1(1+1) |1, m \rangle_{s,\ve{i}}. 
\end{eqnarray}  
As  a   consequence   and  for an  arbitrary set of  $m_{\ve{i}} \in  \left\{ -1,0,1 \right\} $,  
\begin{equation}
\label{gs_degen.eq}
	| \Psi_0      \rangle   =  \otimes_{\ve{i}}  |1,m_{\ve{i}} \rangle_{o,\ve{i}}  
\end{equation}
is  a  ground  state.  Hence  the ground  state  manifold  is  at  least   $3^{N_s}$   degenerate.      

We will now   show  that  the QMC data is consistent,  with the ground state being in the aforementioned manifold  of  states.  In fact,  for 
any  $| \Psi_0    \rangle $  given  by  Eq.~(\ref{gs_degen.eq})   we  have
\begin{eqnarray} 
\label{Eq:Cs}
C^S(\vec{i})=\langle \hat{S}_{\vec{i}}^{1}\cdot \hat{S}_{\vec{0}}^{1}\rangle   = 0. 
\end{eqnarray} 
Note that the  above  holds  for   any  $\alpha$ and our numerical data confirm this point of view.
Our  Hamiltonian  has  a  local orbital    rotational   symmetry, 
\begin{equation}
\label{su2_local.eq}
	   \left[  \hat{H},    \hat{\ve{T}}_{\ve{i}} \right] = 0
\end{equation}
such  that  
\begin{equation}
\label{Eq:CT}
	C^T(\vec{i})=\langle \hat{\ve{T}}_{\vec{i}}\cdot \hat{\ve{T}}_{\vec{0}}\rangle =  \delta_{\ve{i},\ve{0}} 1(1+1).
\end{equation}
In Figs.~\ref{fig:CrKitaev}(a)-(d) we plot the correlators of the   generators of spin and orbital  rotations.  
As  apparent, the QMC  data show $C^S(\vec{i})=0$ as well as $C^T(\vec{i})= 2\delta_{\ve{i},\ve{0}}$,  in accord with the above.

Finally,  we  consider  the dynamical structure factor of the $\hat{\vec{T}}$ generators.
This quantity is defined as  $C^{T}(\boldsymbol{q}, \omega) =  \text{Im} \chi(\boldsymbol{q}, \omega) / \left( 1 - e^{-\beta \omega} \right) $ with 
\begin{eqnarray}
  \label{CSq}
\chi(\boldsymbol{q}, \omega)
&  &=\frac{i}{3} \sum_{\gamma}
\int_0^{\infty} dt  \, 
 e^{i\omega t}
\left<   \left[ \hat{T}^{\gamma}_{\boldsymbol{q}} , \hat{T}^{\gamma}_{\boldsymbol{-q}}(-t) \right] \right>
\end{eqnarray}
where $ \hat{T}^{\alpha}_{\vec{q}}  = \frac{1}{\sqrt{V}} \sum_{\vec{i}} e^{i  \vec{q} \cdot \vec{i}} \left( \hat{T}^{\alpha}_{\vec{i},A}+\hat{T}^{\alpha}_{\vec{i},B}e^{i\vec{q}\vec{R}} \right)$.
Here $\ve{i}$ runs over the $A$ sublattice (or unit cell) on the honeycomb lattice, and $\vec{R}=2/3( \boldsymbol{a_2}-\boldsymbol{a_1}/2)$   with $\boldsymbol{a_1}$ and $\boldsymbol{a_2}$ the  lattice vectors.
We compute this quantity  using the  stochastic  analytical continuation  \cite{Beach04a} as  implemented in the ALF \cite{ALF_v2}  library. 
Our QMC results at $\varphi/\pi=0.5$ and $\varphi/\pi=1.5$ are shown in Figs.~\ref{fig:CrKitaev}(e) and \ref{fig:CrKitaev}(f).
One observes that the spectrum has no momentum dependence and is apparently gapless at any wave vector $\boldsymbol{q}$.
The  absence  of   momentum  dependence stems  from the local SU(2)  invariance   in orbital  space [see Eq.~(\ref{su2_local.eq})].    
We  also  observe  that the  spectral function  does not pick up  excited states.  This stems from the  fact  that the matrix element  
\begin{equation}
	  _{s,\ve{i}}\langle 1, m |     \hat{T}_{\ve{i}}^{\alpha} |1, m'\rangle_{o,\ve{i}}  
\end{equation}
vanishes identically.

\begin{figure}[b]
\centerline{\includegraphics[width=0.35\textwidth]{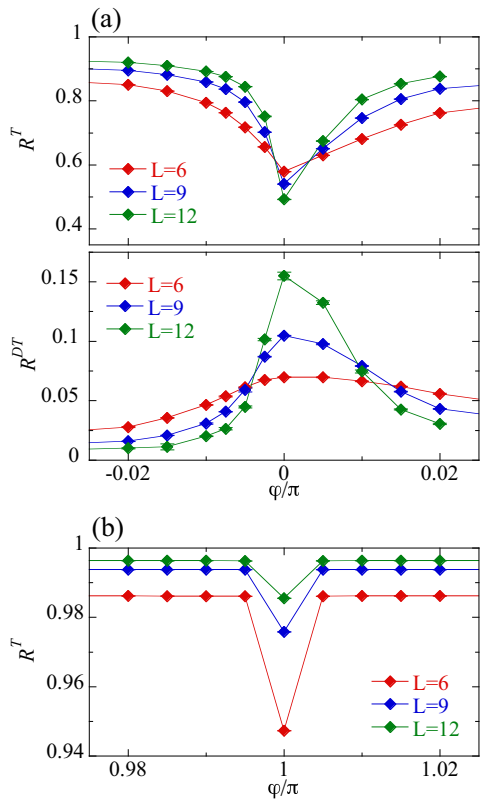}}
\caption{(a) Correlation ratios for AFM (upper panel) and KVBS (down panel) states at different values of $\varphi/\pi$ close to the SU(4) symmetric AFM Heisenberg point at $\varphi/\pi=0$. (b) Correlation ratio for a FM state at different values of $\varphi/\pi$ close to the SU(4) symmetric FM Heisenberg point at $\varphi/\pi=1.0$. Here, $T=1/80$.}
\label{fig:RHA}   
\end{figure}

\begin{figure*}[t]
\centerline{\includegraphics[width=0.95\textwidth]{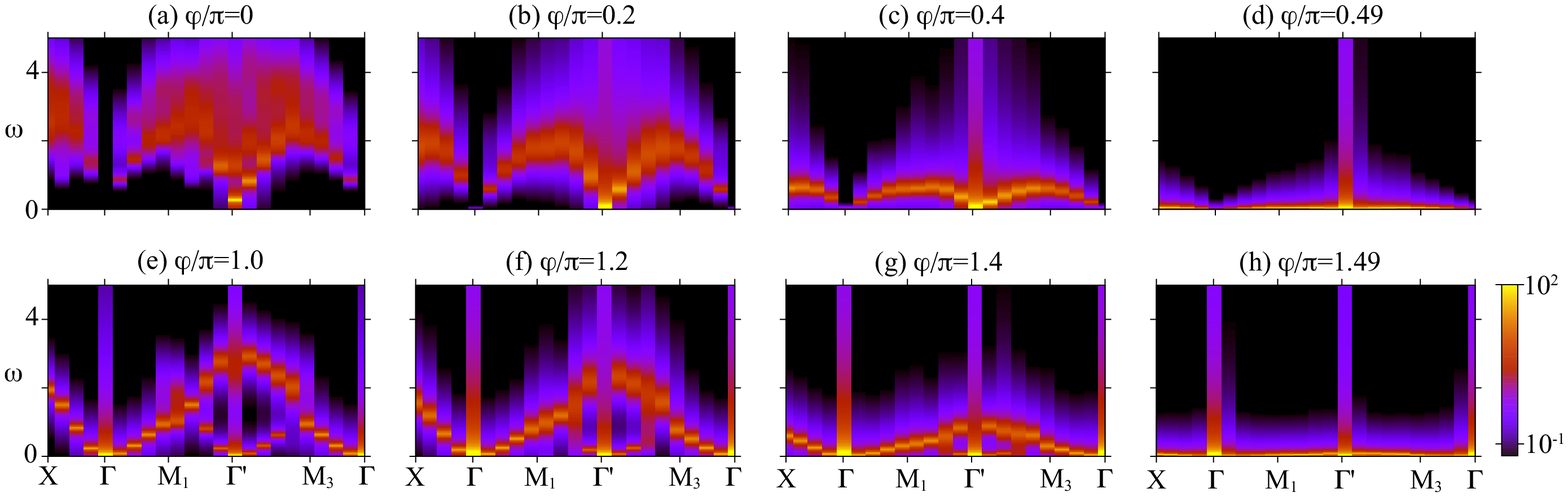}}
\caption{ 
 Dynamical structure factor of $\hat{\vec{T}}$  generators, $C^T(\boldsymbol{q}, \omega)$ at different values of the angle $\varphi$.
 (a), (b), (c), (e), (f), (g) $T=1/80$ and (d), (h) $=1/200$. Results correspond to scans along the red line of first and second Brillouin zones in Fig.~\ref{fig:CrKitaev}(g).
 Here, $L=9$ and $N=2$.}
\label{fig:CqTHeisenberg}   
\end{figure*}

\subsection{From Kitaev to Heisenberg}
As   argued above,  in the Kitaev  limits,  a local SU(2)  symmetry    renders  the  ground  state  macroscopically  degenerate: each  site  as  a three fold  
degeneracy   akin to a  spin-1    orbital  degree  of freedom (see  Eq.~(\ref{eq:S1_orb})).  As  soon as  the  local SU(2)  symmetry is  lifted,  an  S=1   orbital antiferromagnetic  ($\phi/\pi= 0.49$),  Fig.~\ref{fig:Crangle}(c), and  S=1 orbital ferromagnetic  ($\phi/\pi= 1.49$)   Fig.~\ref{fig:Crangle}(f), become  apparent.    This  state is picked  up    by  the  vanishing  of  the   spin correlation function, $C^S(\vec{i})$,  and   ferromagnetic   or  anti-ferromagnetic    ordering  in the orbital degrees of  freedom,  $C^T(\vec{i})$.   Alongside  the  spin and  orbital correlation  functions   we  consider dimer-dimer   ones   that  are  defined  as:  $C^{D^T}(b)=\langle (\hat{{D}}_{b}^{T}-\langle \hat{{D}}_{b}^{T} \rangle)(\hat{{D}}_{0}^{T}-\langle \hat{{D}}_{0}^{T} \rangle)\rangle$. 
For  a  given bond $b = (\ve{i}, \ve{j}) $,   $ \hat{{D}}_{b}^{T}  = \ve{\hat{T}}_{\ve{i}}  \cdot \ve{\hat{T}}_{\ve{j}}   $.    We  note  that  the   dimer-dimer  correlation function  is  a  SU(2)$_o$    singlet  as  it  remains  invariant under SU(2)$_o$    rotations.  At  the Kitaev  points  this   quantity  vanishes   due to  the local   SU(2)$_o$  symmetry.   Away  from  this point  it  is  interesting to see  that  it  shows  substantial    ferromagnetic  correlations.      This  correlation pattern  does  not  break any   further   symmetries  and   merely  reflects  the  ferromagnetic long-ranged  correlations,  Figs.~\ref{fig:Crangle}(d)-(f).

We  note  that  in  a  large  region  around  the    Kitaev points,  the  spin-spin  correlations $C^S(\vec{i})$,    remain  very   small  in  comparison to the orbital ones,  Figs.~\ref{fig:Crangle}(b) and \ref{fig:Crangle}(e).      This  reflects   one of  the key points of our model:  by enhancing the symmetry  from  Z$_2$   to  
Z$_2 \times  SU(2)_o$   one  provides  a means  to  avoid   frustration    via  a  spin-flop  type  transition.    The  spin  flop   transition   is  particularly    visible  in the  vicinity of the SU(4)  ferromagnetic  point,  Fig.~\ref{fig:Crangle}(d).     Here, global  SU(4)  rotations leave  the  
Hamiltonian  invariant,  and one can  find one   that  rotates    $  \hat{\ve{T}}_{\ve{i}} $    to  $  \hat{\ve{S}}_{\ve{i}}$.  As  a  consequence,   both     $C^{S}(\ve{i})$  and  $C^{T}(\ve{i})$    are,  as apparent in Fig.~\ref{fig:Crangle}(d),   identical,  and  show  substantial  ferromagnetic  correlations.  Away  from this  point,   at  $\varphi/\pi = 1.2$ in Fig.~\ref{fig:Crangle}(d),   ferromagnetism is  apparent  only in the orbital   degrees of  freedom.  To  quantify  this spin-flop   transition,  we  consider  the  correlation  ratio.  For  a  general observable  $O$   with  correlations,  
\begin{eqnarray}
C^O_{\gamma  \delta}(\vec{q})=\frac{1}{L^2}\sum_{\ve{R}\ve{R}'}C^O_{\ve{R}\gamma,\ve{R}'\delta}e^{i {\vec q}\cdot (\ve{R}-\ve{R}')}
\label{eq:SF-SM}
\end{eqnarray}
where $\ve{R}, \ve{R}'$ labels the unit cell and $\gamma,\delta$ the orbitals,   it is  given  by 
\begin{eqnarray}
R^{O}=1-\frac{\lambda_1^{O}({\ve q}_0+\delta {\ve q})}{\lambda_1^{O}({\ve q}_0)}
\label{eq:CR-SM}
\end{eqnarray}
Here,    $\lambda_1({\ve q})$ is  the largest eigenvalue  of  the $C^O_{\gamma  \delta}(\vec{q})$  matrix  and $ {\ve q}_0 $  
is  the  ordering  wave  vector    and  ${\ve q}_0 + \delta {\ve q}$   the  largest   wave-length  fluctuation of  the ordered  state  on a  given  lattice  size. 
$R^{O}$  is a   renormalization  group  invariant  quantity~\cite{Binder1981,Pujari16}. 
$R_O\to 1$ for $L\to\infty$ in the ordered state, whereas $R_O\to 0$ in the disordered state.  
At the critical point, $R^O$ is scale-invariant for sufficiently large $L$ so that results for different system sizes cross.
  Fig.~\ref{fig:RHA}(b)     shows  this  quantity  around the   SU(4)   ferromagnetic point.  As  apparent  there  is a  singularity  at 
the SU(4)  point,   that  reflects the   spin-flop  transition.    For  the  considered  totally  antisymmetric  self-conjugate   representation 
of  SU(4),  the   ferromagnetic state   has  a  U(2)$ \times $U(2)    symmetry  whereas   the  Hamiltonian an  U(4)  one.    This  gives   rise  to  a  total  of 
$\text{dim}   \frac{  U(N)} { U(N/2)  \times  U(N/2)}   =  \frac{N^2}{2}  \equiv  n_{BG} $  of  \textit{flat   directions}  of  the fluctuations of the order  parameter or  equivalently  the  number of broken  generators.   As  summarized in Ref.~\cite{Watanabe20},   for  non-relativistic  systems   $n_{BG}$   does not  match   the  number of  Goldstone modes,   $n_{NGM}$.   In fact  $ n_{NGM}   =   n_{BG} -    \frac{1}{2} \text{rank}\left(\rho \right)  $    where  
$ \rho_{\alpha,\beta} =  -\frac{i}{N_s}   \langle  \Psi_0  |  \left[  \hat{T}^{\alpha} , \hat{T}^{\beta}  \right]  | \Psi_0 \rangle $   and $ | \Psi_0 \rangle$ is  the  broken  symmetry  ground  state  and  $  \hat{T}^{\alpha}  =   \sum_{\ve{i}}   \hat{T}_{\ve{i}}^{\alpha} $  the  generators  of  the   global   SU($N$)   symmetry.    Hence,   over  the  spin-flop   transition,  the  number of  Goldstone modes       change  abruptly  from 4  ($N=4$)   to  1   ($N=2$).  Thereby  fluctuations around  the  ordered  state   are  abruptly  suppressed  and  the  ordering  is more  robust.    Magnetically  ordered  states  for the same  representation of  SU($N$)    were   studied  in  Ref.~\cite{Raczkowski20}. 
Finally   we  note  that  away  from the  SU(4)  symmetric  point,   the  correlation   ratio is  next to  angle  independent. 
This behavior is a direct consequence of the fact that  the largest eigenvalue $\lambda_1({\ve q}=\ve{0})$ merely corresponds  to the  total spin,   a  quantity  that  is  determined  by  symmetry  and  not  by Hamiltonian  dynamics.

 We  now  consider  the SU(4) symmetric isotropic AFM Heisenberg realized  at  $\varphi/\pi = 0$.
The ground state of this   model has been reported earlier and corresponds to the KVBS ordered state~\cite{Assaad04,Lang13}.
Our data, Fig.~\ref{fig:Crangle}(a),  supports    this  point of  view    since  the  dimer-dimer  correlation  ratio  grows as  a  function of  system  size 
 whereas the 
AFM one  decreases.     The  KVBS  state  is  a   gaped   state  with  discrete  $C_3$   broken  symmetry  such  that  we  expect  this 
phase  to  be  robust to  perturbations.     As apparent   there is  a   small  window   around  the SU(4)  symmetric point,  where  the data is  consistent with 
the  stability of the  KVBS  phase.       Beyond  the  SU(4)  point,   where  the  model  has  an Z$_2  \times $SU(2)$_o$  symmetry the   data  supports 
 a  continuous  transition  between   the KVBS  and   AFM  phases.     Within   the   theory  of   deconfined   quantum  criticality  \cite{Senthil04_2},   such   an order  to order  transition    requires  the emergence of a  U(1)   spin-liquid  state  at  criticality.  We will see  that  the dynamical   orbital   structure   factor   supports   the  interpretation of  a   two-spinon   continuum  akin  to   such  a 
phase \cite{Assaad16}.      Note  that   the phase boundaries in Fig.~\ref{fig:pd} are based on the crossing points of results for $L=9, 12$.

As  mentioned  above,  we  now  turn our attention to the evolution of the dynamical structure factor of the $\hat{\vec{T}}$ generators, $C^{T}(\boldsymbol{q}, \omega)$.
Fig.~\ref{fig:CqTHeisenberg}  shows typical results at different values of the angle $\varphi$.
At $\varphi/\pi=0$, Fig.~\ref{fig:CqTHeisenberg}(a),    we  are  in  the  KVBS  phase  in the  proximity of  the   deconfined  quantum  critical  point.   The  data  shows  a  small  gap,  and  a  continuum  of  excitations  akin to  the  two-spinon    continuum  of  a  gapless  U(1)  spin liquid.  As  the   angle  
grows    spinons  bind    to  form  a  spin-wave  excitation,  Figs.~\ref{fig:CqTHeisenberg}(b)-(d),    the  velocity of  which  decreases  continuously   and   vanishes  at  the Kitaev  point $\varphi/\pi = 0.5$.

At $\varphi/\pi=1.0$, Fig.~\ref{fig:CqTHeisenberg}(e), the result for $C^{T}(\boldsymbol{q}, \omega)$ produces a well-known quadratic low-lying dispersion around $\vec{\Gamma}$ point.
Moving towards the Kitaev limits [Figs.~\ref{fig:CqTHeisenberg}(f)-(h)]   the  data  again    shows   the reduction of the   spin-wave velocity  in 
the   vicinity of  the  $\vec{\Gamma}$ or $\vec{\Gamma}'$ points.

\section{Summary}
\label{sec:Summary}

Using a phase pinning approach, we have introduced a set of Z$_2\times$ SU($N$)$_o$ generalized Kitaev models that are free of the negative sign  problem for even $N$ within  auxiliary field quantum Monte Carlo simulations.
Our formulation is based on the Abrikosov fermion representation of the spin-1/2 algebra.
The demonstration of the absence of the sign problem stems from the consequence that the imaginary part of the action is quantized to 2$\pi$ by enhancing the number of fermion orbitals from one to $2N$.
This idea allows to provide a generic guideline for defining a set of sign-free models with higher symmetries.   In  fact,  such  a  strategy  was   followed  for  the  Hubbard model in  Ref.~\cite{Assaad02a}  to  stabilize  stripes  in   doped   quantum  anti-ferromagnets. 

We have used this formulation to  investigate  the ground-state properties of the Z$_2\times$SU($2$)$_o$ Kitaev-Heisenberg model.  
Generically,  it is  hard  to predict  the  effect   of  this  symmetry  enhancement  on the  ground state  phase  diagram.  In the  specific  case of  the 
Kitaev model,    it  turns  out  that  the symmetry   enhancement  provides a  route to avoid  frustration.  That is,    at  generic   angles  where  the  symmetry  is not  enhanced,  the  spin-spin  correlations   vanish and   ordering  occurs  in  the  orbital  degrees of  freedom.    
Although the  Kitaev spin liquid phases inherent in original Z$_2$ symmetric model are not present  the ground-state phase diagram 
of  the   symmetry  enhanced model  
is extremely rich.
Aside from the antiferromagnetic and ferromagnetic ordered states with a spontaneously broken SU(2)$_o$ symmetry and the Kekul\'e valence bond solid ordered state with a spontaneously broken translation symmetry, we observe states with higher global SU(4) and local SU(2)$_o$ continuous symmetries.
The  model  equally  supports  a  de-confined  quantum  critical point  between  the  KVBS  and  AFM  with  emergent U(1)   spin liquid  state.

\bigskip
\begin{acknowledgments}
The authors gratefully acknowledge the Gauss Centre for Supercomputing e.V. (www.gauss-centre.eu) for funding this project by providing computing time on the GCS Supercomputer SUPERMUC-NG at the Leibniz Supercomputing Centre (www.lrz.de).
TS thanks funding from the Deutsche Forschungsgemeinschaft under the grant number SA 3986/1-1.
FFA   thanks financial support from the Deutsche Forschungsgemeinschaft,  Project C01 of the  SFB 1170,     as well as the W\"urzburg-Dresden Cluster of Excellence on Complexity and Topology in Quantum Matter ct.qmat (EXC 2147, project-id 390858490). 
\end{acknowledgments}

%\bibliography{./fassaad.bib,./toshihiro.bib,./refs.bib}

\end{document}